\documentclass[runningheads]{llncs}

\usepackage[T1]{fontenc}
\usepackage{graphicx}
\usepackage{booktabs}
\usepackage{amsmath}
\usepackage{amssymb}
\usepackage{array}
\usepackage{url}

\begin{document}

\title{XViewRank: Overlap-Conditioned Score Fusion for LLM-Augmented Sparse Retrieval}
\titlerunning{XViewRank}

\author{Xinxuan Lv
\and Minghan Li\thanks{Corresponding author.}
\and Guodong Zhou}
\authorrunning{X. Lv et al.}
\institute{School of Computer Science and Technology, Soochow University,\\
Suzhou, Jiangsu 215031, China\\
\email{xxlv@stu.suda.edu.cn, mhli@suda.edu.cn, gdzhou@suda.edu.cn}}

\maketitle

\begin{abstract}
Large language models can reduce vocabulary mismatch in sparse retrieval by generating alternative views of a query. Once these views are searched independently, however, their ranked lists may be complementary, redundant, or affected by query drift, making fixed fusion suboptimal. We present XViewRank, a lightweight retrieval-side method for conditionally fusing BM25 results from an original query, a hypothesis view, and a step-back view. For each pooled candidate, XViewRank retains three normalized BM25 scores and three query-level overlaps between the retrieved result sets. A small nonlinear scorer uses the overlaps as context that changes how the view-specific scores are interpreted, without jointly encoding query and passage text. Under a matched-input protocol, XViewRank improves nDCG@10 and MAP numerically over reciprocal rank fusion on both TREC DL 2019 and 2020, with a statistically significant MAP gain on DL19. Additional analyses show that the clearest benefit occurs when the view rankings have low agreement, while the evidence on DL20 is weaker. The results support overlap-conditioned score fusion as a compact retrieval-side alternative to fixed fusion, while also delimiting where its advantage is empirically established.

\end{abstract}

\section{Introduction}

Sparse retrieval remains attractive for large-scale search because inverted indexes are fast, transparent, and easy to deploy \cite{robertson2009probabilistic,lin2021pyserini}. Its weakness is also well known: short user queries often miss the vocabulary and intent cues used in relevant passages \cite{furnas1987vocabulary}. Large language models offer a practical way to reduce this mismatch without rebuilding the index: given an original query, an LLM can generate alternative query views that express the same information need from different angles.

The benefit of multiple query views is not automatic. A hypothesis-style view may introduce useful topical context, while a step-back view may reveal a broader abstraction. At the same time, generated text may drift away from the original intent or duplicate information already captured by another view. After each view is searched with BM25, the system obtains several ranked lists with different score scales and different degrees of overlap \cite{larsen2009data}. Therefore, the retrieval problem is no longer only whether the LLM can generate useful expansion text. A second problem appears after retrieval: how should evidence from several query views be fused while preserving its source and the relationships among the view rankings?

Existing fusion methods address parts of this problem, but leave a gap for interpretable multi-view sparse retrieval. Direct concatenation mixes all query text into one BM25 query, so the system can no longer tell which view produced a candidate. Score-level interpolation keeps view scores separate, but typically uses fixed weights that do not adapt to how much the views overlap for each query. Unsupervised rank fusion is robust to score-scale mismatch, but it treats agreement across views through a fixed rule. Text-based neural reranking can improve final ranking, but it changes the focus from retrieval-side fusion to judging relevance with query and passage text. These choices are useful baselines, but they do not jointly model BM25 evidence from each view and overlap among the view result lists.

We propose XViewRank, a lightweight method for fusing BM25 results from multiple LLM-generated query views. XViewRank keeps the sparse retrieval stage intact. It searches the original query, a hypothesis view, and a step-back view separately, pools the returned documents, and learns a fusion scorer from two interpretable signals. The first signal is BM25 evidence for each candidate from each view. The second signal is how much the view result sets overlap for the current query. XViewRank therefore operates on retrieval-side evidence without requiring a text-based neural reranker. It uses the relationship among the view rankings to condition how candidate-level score evidence is interpreted.

This design keeps evidence from each view visible while allowing different queries to use different fusion patterns. A candidate with similar scores from highly overlapping views should not necessarily be treated the same as a candidate supported by one view when the views are complementary. Because the overlap values are constant across all candidates of one query, an additive linear scorer cannot use them to change the within-query order directly. XViewRank therefore uses a nonlinear scorer, which permits overlap context to change the effective contribution of each view-specific score. Query-balanced training reduces the effect of highly imbalanced sparse-retrieval candidate pools.

We summarize our main contributions:
\begin{itemize}
\item We formulate multi-view sparse retrieval as a matched-input fusion problem. The original, hypothesis, and step-back queries are searched independently, and every matched fusion method receives the same three ranked lists and pooled candidates. This protocol separates improvements due to fusion from improvements due to generating more or better query views.
\item We propose an overlap-conditioned nonlinear scorer that preserves three view-specific BM25 scores and uses three pairwise result-list overlaps as query-level context. The six-dimensional representation allows cross-view agreement to modify the contribution of each retrieval score without joint query and passage encoding or changes to the sparse index.
\item We show that the gain is concentrated in settings where fixed fusion has the greatest difficulty. XViewRank improves nDCG@10 and MAP numerically over matched-view RRF on both TREC DL test years, with significant evidence for DL19 MAP, and its largest pooled gains occur among low-agreement queries. The smaller DL20 gains remain statistically inconclusive.
\end{itemize}

\section{Related Work}

\subsection{LLM Query Views for Sparse Retrieval}

Query expansion has long been used to reduce vocabulary mismatch in sparse retrieval, including classical pseudo-relevance feedback, Rocchio feedback, and relevance models \cite{rocchio1971relevance,lavrenko2001relevance,robertson2009probabilistic}. A recent survey traces how this line has evolved from statistical expansion to pretrained-model and LLM-based generation, and identifies controllability, robustness, and downstream utilization of generated content as continuing challenges \cite{li2026query}. Representative LLM-based methods generate richer query-side text: Query2doc produces pseudo-documents for expansion \cite{wang2023query2doc}; HyDE-style approaches use hypothetical content to bridge the mismatch between queries and documents \cite{gao2023precise}; and hypothesis and step-back prompting expose complementary formulations of the same information need \cite{zhou2024hyqe,zheng2023take}. Dual-layer prompt ensembles further show that combining system- and user-level instructions can improve the robustness of LLM-based expansion and rank fusion \cite{li2026dual}.

Generated retrieval evidence is also used in iterative and domain-specific settings. S2G-RAG explicitly judges evidence sufficiency and remaining information gaps to control iterative retrieval for question answering \cite{li2026s2g}, while GLIER couples generative legal inference with evidence ranking for legal case retrieval \cite{li2026glier}. Related surveys emphasize that long-document retrieval introduces additional challenges in representing and locating evidence across extended contexts \cite{li2025survey}. These studies motivate careful use of generated evidence, but address iterative QA, specialized legal retrieval, or long-document retrieval. XViewRank instead studies a complementary question in general passage retrieval: given several query views generated in advance, how should their sparse retrieval evidence be combined into one ranking?

\subsection{Rank Fusion and Cross-View Evidence}

Combining ranked lists is a classic information retrieval problem. Score fusion methods such as CombSUM combine normalized retrieval scores \cite{fox1994combination,lee1997analyses}, while reciprocal rank fusion is robust to score-scale mismatch \cite{cormack2009reciprocal}. Other probabilistic, segmentation-based, and voting-based fusion methods also aggregate ranked lists with fixed rules \cite{lillis2006probfuse,shokouhi2007segmentation,montague2002condorcet,aslam2001models}. Recent prompt-ensemble retrieval likewise demonstrates that multiple generated formulations can be fused effectively \cite{li2026dual}. However, generated views can be complementary for one query and redundant or drifting for another. XViewRank therefore learns from both view-specific score evidence and query-level view overlap, instead of applying one fixed rule to every query and view combination.

\subsection{Learning to Rank with Lightweight Signals}

Supervised ranking methods learn document order from relevance labels \cite{burges2005learning,cao2007learning,xia2008listwise}. XViewRank follows this direction with retrieval-side features only: it does not encode passage text, but learns a fusion layer from signals that are cheap to compute once BM25 runs are available. Unlike general-purpose supervised fusion models over broad feature sets, XViewRank deliberately restricts the representation to view-specific sparse scores and view-overlap context, so the learned behavior remains tied to the multi-view retrieval problem.

\section{Method}

\begin{figure}[t]
\centering
\includegraphics[width=\linewidth]{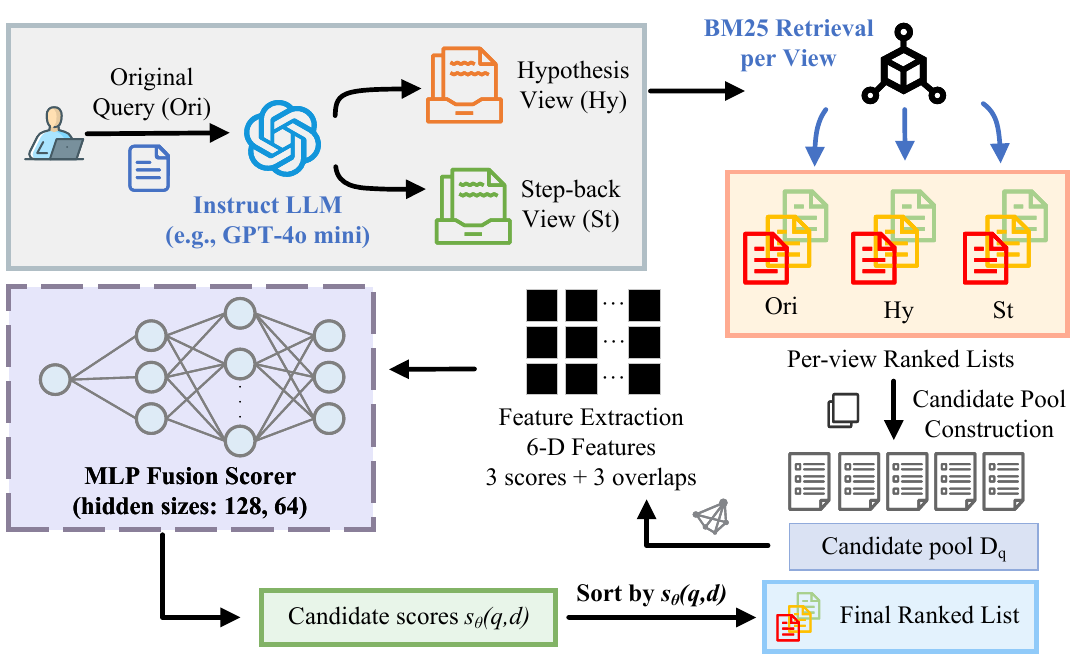}
\caption{Overview of XViewRank. Three query views are searched independently with BM25, merged into a candidate pool, represented by six score-overlap features, and reranked by a lightweight fusion scorer.}
\label{fig:xviewrank-pipeline}
\end{figure}

Figure~\ref{fig:xviewrank-pipeline} gives an overview of XViewRank. The method has three stages: view-specific retrieval, feature construction, and learned fusion.

\subsection{Multi-View Candidate Generation}

For each query \(q\), we construct three query views. The first view is the original query \(q^{ori}\). The second view \(q^{hy}\) is a hypothesis-style rewrite generated by an LLM. The third view \(q^{st}\) is a step-back rewrite that expresses the information need at a more abstract level. We use these two generated views because they represent complementary reformulation directions: the hypothesis view adds concrete topical evidence that may appear in relevant passages, while the step-back view abstracts the information need to recover broader lexical matches. Let \(v \in \{ori,hy,st\}\) denote a view. Each view is searched independently with BM25 over the same passage index, producing a ranked list \(\mathcal{D}_{q}^{v}\). The candidate pool is the de-duplicated union of the documents returned by the three views:
\begin{equation}
\mathcal{D}_q =
\mathcal{D}_{q}^{ori} \cup \mathcal{D}_{q}^{hy} \cup \mathcal{D}_{q}^{st}.
\end{equation}
This design keeps the retrieval evidence from different views separate before fusion. A candidate may receive score evidence from all three views, only from the original query, or only from a generated view.

\subsection{Overlap-Conditioned Score Features}

For a candidate document \(d \in \mathcal{D}_q\), XViewRank represents evidence from two levels: document-level retrieval strength and query-level view relationship.

At the document level, XViewRank records how strongly each query view retrieves the candidate. Because BM25 scores from different generated queries are not directly comparable, scores are normalized within each query-view ranked list using min-max normalization. For a view \(v\), the normalized score is denoted as \(\hat{s}_v(q,d)\). If the maximum and minimum BM25 scores in a view are equal, the normalized score for that view is set to zero. If \(d\) is absent from the ranked list of view \(v\), the corresponding score feature is also set to zero. This yields three normalized score features:
\begin{equation}
\hat{s}_{ori}(q,d),\quad \hat{s}_{hy}(q,d),\quad \hat{s}_{st}(q,d).
\end{equation}
These features keep the evidence from the original query, the hypothesis view, and the step-back view separate.

At the query level, XViewRank records how much the views agree with each other. Let \(\mathcal{D}_{q,v}^{\tau}\) be the top-\(\tau\) documents retrieved by view \(v\). For each pair of views \(u\) and \(v\), we compute the Jaccard overlap between their top-\(\tau\) result sets:
\begin{equation}
J_{u,v}(q)=
\frac{|\mathcal{D}_{q,u}^{\tau} \cap \mathcal{D}_{q,v}^{\tau}|}
{|\mathcal{D}_{q,u}^{\tau} \cup \mathcal{D}_{q,v}^{\tau}|}.
\end{equation}
We use \(\tau=500\) in all experiments. The three overlap features \(J_{ori,hy}(q)\), \(J_{ori,st}(q)\), and \(J_{hy,st}(q)\) describe whether the generated views are mostly redundant with the original query or provide complementary evidence.

The final feature vector for candidate \(d\) is:
\begin{equation}
x(q,d)=
[\hat{s}_{ori},\hat{s}_{hy},\hat{s}_{st},
J_{ori,hy},J_{ori,st},J_{hy,st}].
\end{equation}
This representation is intentionally small and transparent. The score features indicate whether a candidate is strongly supported by each view, while the overlap features provide the query-level context needed to interpret that support. The overlap values are constant for all candidates of the same query, but they can still affect within-query ranking through the nonlinear scorer: the same score pattern can be mapped differently when the views are highly redundant than when they retrieve mostly complementary result sets.

More formally, let \(\mathbf{s}(q,d)\) contain the three document-level score features and let \(\mathbf{o}(q)\) contain the three query-level overlap features. The scorer has the conditional form
\begin{equation}
s_{\theta}(q,d)=f_{\theta}(\mathbf{s}(q,d),\mathbf{o}(q)).
\end{equation}
Although \(\mathbf{o}(q)\) is constant within a query, the nonlinear scorer allows it to change the local sensitivity \(\partial s_{\theta}(q,d)/\partial \hat{s}_{v}(q,d)\) to each view-specific score. Thus, overlap is not treated as direct document relevance evidence. It conditions how the model interprets candidate-level retrieval evidence. Section~4.5 verifies this behavior through linear controls, overlap perturbations, and input-gradient analysis.

\subsection{Fusion Scorer}

XViewRank uses a small multilayer perceptron to map the score-overlap feature vector to a relevance score:
\begin{equation}
s_{\theta}(q,d)=f_{\theta}(x(q,d)).
\end{equation}
The model has two hidden layers with 128 and 64 hidden units, ReLU activations, and dropout rate 0.2. It is intentionally much smaller than a text-based reranker. The goal is not to model full semantic matching from passage text, but to combine retrieval evidence across views.

\subsection{Training Objective}

The fusion scorer is trained with weighted query-balanced binary cross-entropy. During training, \(\mathcal{D}_q\) is the union of the top-1,000 BM25 results from the three views. A candidate is positive if it has a positive MS MARCO relevance label for the query, and unlabeled candidates in the pool are treated as negative. Let \(y_{q,d}\in\{0,1\}\) denote this label. Since sparse retrieval candidate pools contain far more negatives than positives, we use a capped positive-class weight. Let \(P\) and \(N\) be the total numbers of positive and negative training candidates in the training pool. The positive-class weight is \(\gamma=\min(N/P,200)\). For a positive candidate, we multiply \(\gamma\) by \(1/\log_2(r_{\min}+2)\), where \(r_{\min}\) is the best rank of the candidate among the three view-specific ranked lists; negative candidates receive weight 1. Let \(w_{q,d}\) denote the resulting example weight. The loss is:
\begin{equation}
\mathcal{L}
=
\frac{1}{|\mathcal{Q}|}
\sum_{q \in \mathcal{Q}}
\frac{1}{|\mathcal{D}_q|}
\sum_{d \in \mathcal{D}_q}
w_{q,d}\,\mathrm{BCE}(y_{q,d}, s_{\theta}(q,d)).
\end{equation}
Here, \(s_{\theta}(q,d)\) is a logit, and \(\mathrm{BCE}(y,s)=-y\log\sigma(s)-(1-y)\log(1-\sigma(s))\), where \(\sigma\) is the sigmoid function. This objective trains a cross-view relevance scorer while preventing queries with larger candidate pools from dominating training. The default XViewRank configuration uses this pointwise objective because it is simple, stable across TREC DL years, and does not require test-year-specific model selection.

\section{Experiments}

\subsection{Datasets and Evaluation}

We evaluate on the MS MARCO passage collection with TREC DL 2019 and TREC DL 2020 topics \cite{craswell2020overview,craswell2021overview}. Retrieval is performed with Pyserini BM25 \cite{lin2021pyserini}. Following our experimental protocol, BM25 uses \(k_1=0.9\) and \(b=0.4\). All methods return up to 1,000 documents per query. We report nDCG@10, MAP, and Recall@1000. For MAP and Recall@1000, relevance is binarized at grade 2 for TREC DL, following the strict relevance setting commonly used for binary passage-retrieval metrics. We also include a zero-shot transfer check on FiQA and NFCorpus, where the MS MARCO-trained scorer is applied without dataset-specific retraining.

XViewRank is trained with MS MARCO passage training relevance labels and selected on a 2,000-query MS MARCO development sample. After removing training queries without a positive pooled candidate and excluding development-query overlap, the final training pool contains 6,317 queries and 9,040,208 query-document candidates, including 6,602 positives. Both generated views are available for every retained training query. The development sample has no query overlap with the TREC DL 2019 or 2020 topics. TREC DL 2019 and 2020 are used only as held-out test sets, and their qrels are not used for model selection. Hyperparameters and model checkpoints are selected using the development set, and the same selected configuration is then applied to both test years.

\subsection{LLM Views and Implementation Details}

All LLM-generated query views and LLM expansion baselines use GPT-4o mini. For XViewRank, the hypothesis and step-back views are generated once, cached, and reused by every method requiring the same view text. The XViewRank scorer uses the score-overlap representation described in Section~3.2 and is selected exclusively on the 2,000-query development split; neither DL19 nor DL20 qrels are used for model or hyperparameter selection. The exact prompts, decoding parameters, retrieval settings, training configuration, and evaluation conventions are reported in Appendix~\ref{app:reproducibility}.

The main table reports one fixed development-selected configuration applied unchanged to both test years. Sensitivity and view-portfolio experiments retrain the scorer for each analyzed configuration; their absolute values may therefore differ slightly from the fixed main run and should be interpreted as within-table comparisons.

\subsection{Baselines}

We compare with three groups of baselines. The first group contains sparse retrieval and classical query expansion: BM25, BM25 with RM3, BM25 with Rocchio-style feedback, and BM25 over a doc2query-T5 expanded Lucene index~\cite{nogueira2019doc2query}. The doc2query-T5 baseline represents document-side expansion, while XViewRank keeps the sparse index unchanged and learns to combine query-side retrieval views. The second group contains LLM expansion baselines. We include Zero-shot pseudo-doc, a pseudo-document expansion baseline inspired by Query2doc~\cite{wang2023query2doc}; unlike the original few-shot Query2doc setting, it uses no in-context demonstrations. We also include Zero-shot CoT, a chain-of-thought-style query expansion baseline following prompt-based query expansion~\cite{jagerman2023query}. Finally, we compare with the original Exp4Fuse setting, which fuses the original query view with the hypothesis view. The third group contains fixed fusion baselines over the same matched view set used by XViewRank. In particular, RRF over matched views uses the original view, the hypothesis view, and the same step-back view as XViewRank; therefore its comparison isolates fixed fusion from learned fusion under identical inputs. We also include a dev-tuned weighted CombSUM baseline, where nonnegative weights for the three normalized view scores are selected on the development set and then fixed for DL19 and DL20.

This setup separates two questions. The original Exp4Fuse baseline measures the established two-view LLM expansion setting. The matched-view fusion baselines measure whether the same input views are better combined by fixed rules or by the learned XViewRank scorer.

\subsection{Main Results}

Table~\ref{tab:main-results} reports the main results. Exp4Fuse uses the original and hypothesis views, following its published two-view setting. CombSUM, weighted CombSUM, RRF, and XViewRank all use the same original, hypothesis, and step-back views; these rows therefore provide the matched-input fusion comparison. XViewRank obtains the highest nDCG@10 and MAP among the reported methods on both TREC DL 2019 and 2020. Compared with BM25, classical feedback, and document-side expansion with doc2query-T5, the gains show that generated query views provide useful additional retrieval evidence when their scores are fused carefully. Compared with fixed matched-view fusion methods, including RRF and dev-tuned weighted CombSUM, XViewRank gives higher nDCG@10 and MAP while keeping Recall@1000 nearly unchanged. We test the main comparison against RRF with paired sign-flip tests over queries and use bootstrap confidence intervals as an additional check; because TREC DL contains few topics, we interpret significance conservatively. This pattern suggests that the learned scorer mainly improves head ranking by combining cross-view score evidence more effectively, rather than simply expanding the candidate pool.

\begin{table}[t]
\centering
\caption{Main results on TREC DL passage retrieval. The doc2query-T5 baseline uses a document-expanded Lucene index; the remaining sparse and multi-view runs use the MS MARCO passage index. All runs return up to 1,000 documents per query.}
\label{tab:main-results}
\scriptsize
\setlength{\tabcolsep}{1.6pt}
\begin{tabular*}{\linewidth}{@{\extracolsep{\fill}}lcccccc}
\toprule
Method & \multicolumn{3}{c}{DL19} & \multicolumn{3}{c}{DL20} \\
\cmidrule(lr){2-4}\cmidrule(lr){5-7}
 & nDCG@10 & MAP & R@1000 & nDCG@10 & MAP & R@1000 \\
\midrule
BM25 & 0.5058 & 0.3013 & 0.7501 & 0.4796 & 0.2856 & 0.7863 \\
BM25+RM3 & 0.5216 & 0.3416 & 0.8136 & 0.4896 & 0.3006 & 0.8236 \\
BM25+Rocchio & 0.5275 & 0.3474 & 0.8007 & 0.4910 & 0.3115 & 0.8156 \\
doc2query-T5 BM25 & 0.6417 & 0.4034 & 0.8310 & 0.6187 & 0.4074 & 0.8452 \\
Zero-shot pseudo-doc & 0.5812 & 0.3833 & 0.8549 & 0.5477 & 0.3720 & 0.8688 \\
Zero-shot CoT & 0.6330 & 0.4086 & 0.8740 & 0.6099 & 0.4046 & 0.8677 \\
Exp4Fuse~\cite{liu2025exp4fuse} & 0.6199 & 0.3853 & 0.8698 & 0.6123 & 0.4122 & 0.8844 \\
CombSUM & 0.6134 & 0.4183 & 0.8622 & 0.5947 & 0.4420 & 0.8887 \\
Weighted CombSUM & 0.6346 & 0.4055 & {\bfseries 0.8897} & 0.6192 & 0.4246 & {\bfseries 0.9034} \\
RRF & 0.6376 & 0.4225 & 0.8893 & 0.6152 & 0.4399 & 0.9018 \\
XViewRank & {\bfseries 0.6583} & {\bfseries 0.4437}\(\dagger\) & 0.8893 & {\bfseries 0.6250} & {\bfseries 0.4480} & 0.9013 \\
\bottomrule
\end{tabular*}
\vspace{1mm}
\footnotesize
Exp4Fuse follows its original two-view configuration and is not the matched-input control. \(\dagger\) indicates significant improvement over matched-view RRF under a paired sign-flip test (\(p<0.05\)). Bold numbers indicate the best result in each column.
\end{table}

Table~\ref{tab:rrf-significance} reports a paired comparison with the standard matched-view RRF baseline. XViewRank improves nDCG@10 and MAP numerically on both test years, but the evidence is not uniformly significant: DL19 MAP reaches \(p<0.05\), whereas DL19 nDCG@10 and the smaller DL20 gains do not. We therefore do not claim universal superiority. The supported conclusion is that overlap-conditioned score fusion improves the shared candidate ordering most clearly on DL19, while the agreement-stratified analysis below identifies the subset of queries on which this advantage is concentrated.

\begin{table}[t]
\centering
\caption{Paired significance of XViewRank over matched-view RRF. Differences are XViewRank minus RRF. Per-query values use the same trec\_eval-compatible metric definitions as Table~\ref{tab:main-results}. Sign-flip tests use 100,000 permutations; confidence intervals are obtained from 10,000 query bootstrap samples.}
\label{tab:rrf-significance}
\scriptsize
\setlength{\tabcolsep}{3pt}
\begin{tabular}{llrrl}
\toprule
Dataset & Metric & Difference & \(p\)-value & Bootstrap 95\% CI \\
\midrule
DL19 & nDCG@10 & +0.0207 & 0.0672 & [0.0001, 0.0434] \\
DL19 & MAP & +0.0212 & \textbf{0.0457} & [0.0021, 0.0467] \\
DL19 & R@1000 & +0.0001 & 0.9736 & [-0.0047, 0.0053] \\
DL20 & nDCG@10 & +0.0098 & 0.2684 & [-0.0069, 0.0271] \\
DL20 & MAP & +0.0081 & 0.2498 & [-0.0041, 0.0221] \\
DL20 & R@1000 & -0.0005 & 0.8316 & [-0.0046, 0.0030] \\
\bottomrule
\end{tabular}
\end{table}

\subsection{Ablations and Analysis}

We analyze XViewRank through several questions. Do the generated views provide useful but incomplete evidence? Do score and overlap signals explain the learned fusion behavior? Is the method sensitive to the overlap depth \(\tau\)? Does the model really use overlap context? Is the lightweight nonlinear scorer necessary? Finally, does XViewRank improve ranking quality beyond a fixed fusion rule when the input views are held constant?

\subsubsection{Are generated views complementary?}
Table~\ref{tab:view-portfolio} compares different view portfolios. Each generated view alone improves over the original query on nDCG@10, but the strongest results come from combining views. No two-view portfolio reaches the all-view setting. This suggests that the hypothesis and step-back views provide complementary retrieval evidence, and that XViewRank benefits from combining all three views rather than relying on a single generated rewrite.

\begin{table}[t]
\centering
\caption{View portfolio ablation. Values report nDCG@10 and MAP.}
\label{tab:view-portfolio}
\small
\setlength{\tabcolsep}{4pt}
\begin{tabular}{lcccc}
\toprule
Portfolio & \multicolumn{2}{c}{DL19} & \multicolumn{2}{c}{DL20} \\
\cmidrule(lr){2-3}\cmidrule(lr){4-5}
 & nDCG@10 & MAP & nDCG@10 & MAP \\
\midrule
Original only & 0.5058 & 0.3013 & 0.4796 & 0.2856 \\
Hypothesis only & 0.5892 & 0.3862 & 0.5293 & 0.3788 \\
Step-back only & 0.5729 & 0.3872 & 0.5833 & 0.4340 \\
Original + hypothesis & 0.6077 & 0.3835 & 0.6051 & 0.4015 \\
Original + step-back & 0.6148 & 0.3878 & 0.5778 & 0.3727 \\
Hypothesis + step-back & 0.6011 & 0.4107 & 0.5854 & 0.4279 \\
All three views & {\bfseries 0.6558} & {\bfseries 0.4406} & {\bfseries 0.6214} & {\bfseries 0.4445} \\
\bottomrule
\end{tabular}
\end{table}

\subsubsection{Do score and overlap signals matter?}

Table~\ref{tab:ablation} validates the two signals used by XViewRank. The score-only variant uses only the three normalized BM25 score features. The document-features-only variant uses document-level view indicators such as rank, view presence, and view count, but removes the query-level view-overlap features. The extended-features variant combines those document-level indicators with overlap. The results show that score features alone are not sufficient, and removing overlap causes a large drop, especially on DL19. Adding broader document-level indicators also does not consistently improve over the score-overlap representation. This supports the central design choice: the model should know not only how strongly each view retrieves a candidate, but also how redundant or complementary the views are for the query.

The particularly low DL20 result of the document-features-only variant deserves clarification. Rank, presence, and view count describe where a document occurs, but they discard the magnitude of its normalized retrieval support. Their distributions are also more sensitive to changes in candidate-pool composition across test years. The result therefore does not imply that provenance indicators are intrinsically harmful; it shows that they are a weak substitute for view-specific score evidence and transfer less reliably on their own.

\begin{table}[t]
\centering
\caption{Feature signal analysis. Values report the same learning protocol with different retrieval-side signals.}
\label{tab:ablation}
\small
\setlength{\tabcolsep}{4pt}
\begin{tabular}{lcccc}
\toprule
Variant & \multicolumn{2}{c}{DL19} & \multicolumn{2}{c}{DL20} \\
\cmidrule(lr){2-3}\cmidrule(lr){4-5}
 & nDCG@10 & MAP & nDCG@10 & MAP \\
\midrule
Score only & 0.6045 & 0.3776 & 0.6232 & 0.4076 \\
Document features only & 0.6189 & 0.3878 & 0.5862 & 0.3988 \\
Extended features with overlap & 0.6419 & 0.4305 & 0.6218 & 0.4467 \\
XViewRank & {\bfseries 0.6583} & {\bfseries 0.4437} & {\bfseries 0.6250} & {\bfseries 0.4480} \\
\bottomrule
\end{tabular}
\end{table}

\subsubsection{Does the model use the overlap values?}

To test whether the association between a query and its overlap context matters beyond the marginal overlap distribution, we perturb only the three overlap features at inference time while keeping the view-specific BM25 scores and trained checkpoint fixed. We independently permute complete overlap vectors among queries 100 times, preserving both the marginal distribution and the dependence among the three overlap dimensions while breaking their association with the current query. Table~\ref{tab:overlap-perturb} reports the change from the unpermuted control to the mean of the permuted runs. Empirical one-sided \(p\)-values are computed as \((1+n_{\mathrm{shuffle}\geq\mathrm{real}})/(100+1)\). Shuffling reduces both nDCG@10 and MAP on DL19, whereas the changes on DL20 are small and statistically insignificant. Thus, query-specific overlap context is supported on DL19; the same mechanism is not statistically established on DL20. The perturbation code reconstructs both the control and shuffled rankings through one dedicated inference path. We therefore report paired changes within that path rather than comparing its absolute scores with the separately archived main run in Table~\ref{tab:main-results}.

\begin{table}[t]
\centering
\caption{Repeated overlap permutation analysis with 100 independent query-level shuffles. Changes are the shuffled mean minus the unpermuted control, so negative values indicate degradation; \(p\) is the empirical one-sided permutation probability.}
\label{tab:overlap-perturb}
\scriptsize
\setlength{\tabcolsep}{2.5pt}
\begin{tabular}{lrrrr}
\toprule
Dataset & \multicolumn{2}{c}{Change after shuffling} & \multicolumn{2}{c}{Empirical \(p\)} \\
\cmidrule(lr){2-3}\cmidrule(lr){4-5}
 & nDCG@10 & MAP & nDCG@10 & MAP \\
\midrule
DL19 & -0.0098 & -0.0134 & \textbf{0.0198} & \textbf{0.0099} \\
DL20 & +0.0015 & -0.0013 & 0.6634 & 0.3564 \\
\bottomrule
\end{tabular}
\end{table}

\subsubsection{Is the overlap depth important?}

Table~\ref{tab:tau-sensitivity} varies the overlap depth used to compute the Jaccard features. The results are stable from \(\tau=50\) to \(\tau=1000\), and \(\tau=500\) is not a fragile choice. This also indicates that the overlap signal captures a broad property of view agreement rather than a narrow artifact of one cutoff.

\begin{table}[t]
\centering
\caption{Sensitivity to the overlap depth \(\tau\). Values report nDCG@10 and MAP.}
\label{tab:tau-sensitivity}
\small
\setlength{\tabcolsep}{5pt}
\begin{tabular}{rcccc}
\toprule
\(\tau\) & \multicolumn{2}{c}{DL19} & \multicolumn{2}{c}{DL20} \\
\cmidrule(lr){2-3}\cmidrule(lr){4-5}
 & nDCG@10 & MAP & nDCG@10 & MAP \\
\midrule
50 & 0.6589 & 0.4373 & 0.6265 & 0.4453 \\
100 & 0.6569 & 0.4376 & {\bfseries 0.6290} & 0.4471 \\
200 & 0.6522 & 0.4390 & 0.6268 & {\bfseries 0.4490} \\
500 & 0.6558 & 0.4386 & 0.6243 & 0.4459 \\
1000 & 0.6552 & {\bfseries 0.4393} & 0.6226 & 0.4431 \\
\bottomrule
\end{tabular}
\end{table}

\subsubsection{Why use a small MLP scorer?}

Table~\ref{tab:fusion-scorer} compares learned fusion models using the same matched view runs. Linear scores uses only the three normalized BM25 scores. Linear scores+overlap adds the three query-level overlap constants. These additive constants cannot directly change document order within a query; any difference from Linear scores arises because retraining changes the learned coefficients on the score features. Linear overlap weights instead adds explicit products between scores and overlaps, which provides a direct linear interaction but proves unstable in this experiment. We also report a LambdaMART control using the same six raw features, with 15 leaves, learning rate 0.03, at most 800 trees, and development-set early stopping. The last row is the default XViewRank scorer from Table~\ref{tab:main-results}. The MLP performs best, consistent with the need for nonlinear interactions between document-level scores and query-level overlap context. LambdaMART is a matched-feature control rather than an exhaustive hyperparameter study.

\begin{table}[t]
\centering
\caption{Fusion scorer comparison under the same three-view inputs. Values report nDCG@10 and MAP.}
\label{tab:fusion-scorer}
\scriptsize
\setlength{\tabcolsep}{3pt}
\begin{tabular}{lcccc}
\toprule
Scorer & \multicolumn{2}{c}{DL19} & \multicolumn{2}{c}{DL20} \\
\cmidrule(lr){2-3}\cmidrule(lr){4-5}
 & nDCG@10 & MAP & nDCG@10 & MAP \\
\midrule
Linear scores & 0.6146 & 0.4201 & 0.5932 & 0.4322 \\
Linear scores+overlap & 0.6267 & 0.4218 & 0.6004 & 0.4368 \\
Linear overlap weights & 0.3747 & 0.2064 & 0.4264 & 0.2690 \\
LambdaMART (same features) & 0.5530 & 0.3543 & 0.5432 & 0.3466 \\
Score-overlap MLP & {\bfseries 0.6583} & {\bfseries 0.4437} & {\bfseries 0.6250} & {\bfseries 0.4480} \\
\bottomrule
\end{tabular}
\end{table}

\subsubsection{Does learned fusion improve over fixed fusion?}

The most direct fixed-fusion comparison is RRF over the same original, hypothesis, and step-back views. XViewRank improves over this baseline by 0.0207 nDCG@10 and 0.0212 MAP on DL19. The MAP gain reaches significance under a paired sign-flip test (\(p=0.0457\)), while the nDCG@10 sign-flip \(p\)-value remains above 0.05 despite a positive bootstrap interval. On DL20, the gains are 0.0098 nDCG@10 and 0.0081 MAP and are not significant. Recall@1000 remains nearly unchanged on both years. Therefore, the advantage of XViewRank comes mainly from reordering the shared candidate pool more effectively, rather than from retrieving many additional tail documents. Given the small number of topics and the several reported metrics, we treat the DL19 MAP result as limited evidence rather than a broad claim of statistical superiority.

The scorer comparison in Table~\ref{tab:fusion-scorer} provides another matched-view check. Together with the CombSUM and weighted CombSUM results in Table~\ref{tab:main-results}, it suggests that the gain is not merely due to pooling the three views, but to learning a nonlinear combination of score evidence under view-overlap context.

\subsubsection{How do query-level overlap features affect ranking?}

The overlap features are constant for all candidates of the same query. A plain linear model without interaction terms cannot use such constants to change the within-query order, but a nonlinear scorer can use them to change how score features are interpreted. We check this behavior by computing input-gradient magnitudes with respect to the three score features and grouping queries by average view overlap. Table~\ref{tab:overlap-gradient} shows that total score sensitivity is higher for low- and medium-overlap queries than for high-overlap queries, while generated-view scores account for most of the gradient mass across bins. This supports the intended role of overlap features as query-level context for score combination.

\begin{table}[t]
\centering
\caption{Gradient-based analysis of score-feature sensitivity. Total gradient is the sum of absolute gradients over the three score features; generated-view share is the normalized gradient mass on the hypothesis and step-back score features.}
\label{tab:overlap-gradient}
\scriptsize
\setlength{\tabcolsep}{2.5pt}
\begin{tabular}{lcccccc}
\toprule
Bin & \(O_{19}\) & Grad19 & Gen19 & \(O_{20}\) & Grad20 & Gen20 \\
\midrule
Low & 0.1597 & 3.9307 & 0.7179 & 0.1148 & 4.1741 & 0.7153 \\
Medium & 0.2592 & 3.5884 & 0.7283 & 0.2163 & 4.1547 & 0.7117 \\
High & 0.4592 & 3.0937 & 0.7510 & 0.3215 & 3.5872 & 0.7329 \\
\bottomrule
\end{tabular}
\end{table}

\begin{figure}[t]
\centering
\includegraphics[width=0.98\linewidth]{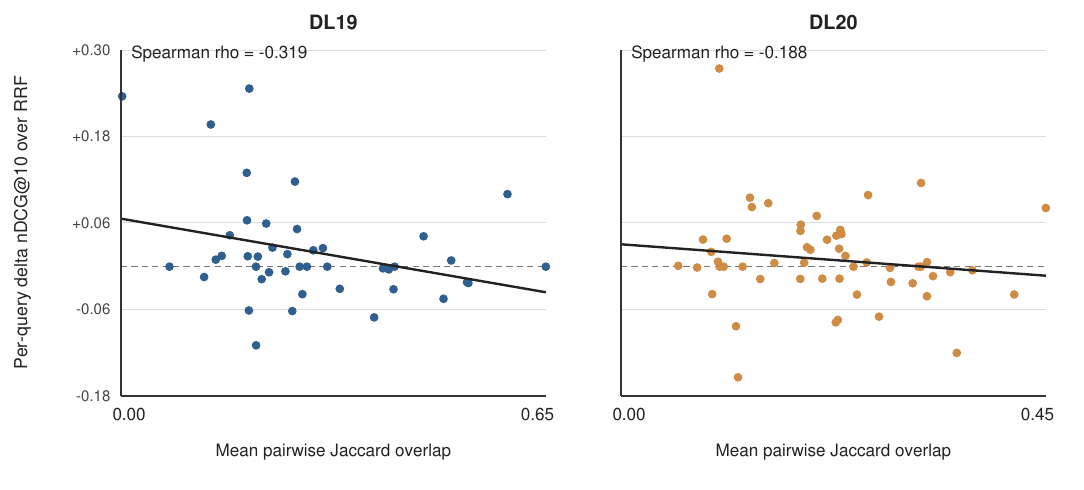}
\caption{Per-query relationship between average pairwise top-500 Jaccard overlap and the nDCG@10 gain of XViewRank over matched-view RRF. The solid line is a linear trend shown for visualization; inference uses Spearman correlation with a permutation test.}
\label{fig:overlap-gain-continuous}
\end{figure}

Figure~\ref{fig:overlap-gain-continuous} examines the relationship without relying on discrete bins. On DL19, average pairwise overlap is negatively associated with the per-query nDCG@10 gain over RRF (Spearman \(\rho=-0.3190\), permutation \(p=0.0353\)). DL20 has the same direction but a weaker, non-significant association (\(\rho=-0.1878\), \(p=0.1758\)). The continuous analysis therefore supports the low-agreement mechanism on DL19 while showing that its strength varies across test years.

Table~\ref{tab:agreement-bins} quantifies this pattern by computing the mean of the three pairwise top-100 Jaccard overlaps, forming tertiles separately within DL19 and DL20, and then pooling corresponding bins. The top-100 cutoff focuses the analysis on the head of the rankings and is distinct from the default top-500 feature depth used by XViewRank. Accordingly, this stratified result is a complementary head-focused diagnostic rather than direct evidence about the exact default overlap feature. The low-agreement group shows significant gains for both nDCG@10 and MAP, whereas the medium- and high-agreement groups do not. This result narrows the mechanism claim: XViewRank is most useful when the retrieval views expose different candidate rankings that fixed rank fusion cannot resolve as effectively. Low agreement is an observed condition under which learned fusion has more room to help; it should not be interpreted as a generation objective, because deliberately forcing unrelated views can instead introduce query drift.

\begin{table}[t]
\centering
\caption{Matched-view RRF and XViewRank by pooled top-100 Jaccard agreement tertile. Tertiles are formed within each test split before pooling. Bold \(p\)-values are significant under paired sign-flip tests.}
\label{tab:agreement-bins}
\scriptsize
\setlength{\tabcolsep}{2.5pt}
\begin{tabular}{llrrrrr}
\toprule
Bin & Metric & Queries & RRF & XViewRank & Difference & \(p\)-value \\
\midrule
Low & nDCG@10 & 32 & 0.5572 & 0.5995 & +0.0422 & \textbf{0.0059} \\
Low & MAP & 32 & 0.3998 & 0.4336 & +0.0338 & \textbf{0.0025} \\
Medium & nDCG@10 & 33 & 0.6161 & 0.6254 & +0.0093 & 0.2497 \\
Medium & MAP & 33 & 0.3970 & 0.3981 & +0.0012 & 0.8162 \\
High & nDCG@10 & 32 & 0.7024 & 0.6950 & -0.0074 & 0.4689 \\
High & MAP & 32 & 0.5009 & 0.5080 & +0.0071 & 0.5448 \\
\bottomrule
\end{tabular}
\end{table}

\begin{table}[t]
\centering
\caption{Zero-shot transfer to two BEIR datasets. RRF, CombSUM, and XViewRank use the matched three-view inputs; Exp4Fuse uses its original two-view input. Values report nDCG@10.}
\label{tab:beir-transfer}
\small
\setlength{\tabcolsep}{5pt}
\begin{tabular}{lcccc}
\toprule
Dataset & CombSUM & RRF & Exp4Fuse-OH & XViewRank \\
\midrule
FiQA & 0.2275 & 0.2389 & 0.2305 & {\bfseries 0.2482} \\
NFCorpus & 0.3464 & 0.3456 & 0.3503 & {\bfseries 0.3510} \\
\bottomrule
\end{tabular}
\end{table}

\subsubsection{Does the learned scorer transfer to other collections?}

We further test zero-shot transfer to two BEIR datasets. The XViewRank scorer is still trained on MS MARCO and is directly applied to FiQA and NFCorpus without dataset-specific tuning. RRF, CombSUM, and XViewRank use the same original, hypothesis, and step-back views for each target dataset, while Exp4Fuse follows its original two-view setting with the original and hypothesis views. Table~\ref{tab:beir-transfer} reports nDCG@10. XViewRank obtains the highest score among the reported baselines on both datasets. We treat this as a limited transfer check rather than a claim of universal cross-collection robustness.

XViewRank is a retrieval-side fusion method and does not jointly encode query and passage text. A matched-budget neural-reranking reference is therefore reported separately in Appendix~\ref{app:neural-reference}, rather than being treated as part of the main method.

\section{Discussion}

\paragraph{What is new in XViewRank?}
The individual ingredients of XViewRank, namely BM25 scores, Jaccard overlap, and a small MLP, are established. The contribution lies in their use as a matched-view conditional fusion formulation for LLM-generated sparse retrieval. Unlike concatenation, XViewRank preserves each score together with the query view that produced it. Unlike fixed RRF or CombSUM, its nonlinear scorer lets query-level agreement change how view-specific score evidence is interpreted. The matched-input protocol is essential because it separates the effect of fusion from the quality or number of generated views.

\paragraph{When does overlap help?}
The repeated perturbation and gradient analyses indicate that overlap acts as coarse context rather than direct relevance evidence. Agreement-stratified and continuous analyses further show that the improvement is concentrated in low-agreement queries on DL19, where the views expose heterogeneous candidate rankings; DL20 follows the same broad direction but does not provide significant continuous or permutation evidence. This observation does not imply that arbitrary diversity is beneficial. Low agreement can arise from useful complementary formulations or from harmful query drift; XViewRank only receives retrieval-side evidence and cannot fully distinguish these causes.

\paragraph{Deployment scope.}
The LLM generation cost is shared with the LLM-expansion baselines and can be amortized when views are cached. After the three sparse runs are available, XViewRank adds only a six-feature MLP evaluation per pooled candidate and does not require joint query and passage text encoding. Neural rerankers remain complementary when additional latency is acceptable, as illustrated in Appendix~\ref{app:neural-reference}. Accordingly, the main comparison in this work is against fixed and learned retrieval-side fusion under identical views, while the cross-encoder experiment serves only to locate XViewRank within a multi-stage pipeline.

\section{Limitations}

The study has several boundaries. First, DL19 and DL20 contain relatively few deeply judged topics, which limits the power of query-level significance tests. Only the DL19 MAP improvement over matched-view RRF is significant in the full-split comparison, and the DL20 gains should be interpreted as numerical rather than statistically established. The repeated permutation and continuous correlation analyses likewise support the overlap mechanism on DL19 but not on DL20. Second, the main controlled experiments use query views from one LLM configuration. Caching makes fusion comparisons fair, but does not establish robustness to a different generator or prompt distribution. Third, XViewRank requires labeled source-domain data to train the fusion scorer. The FiQA and NFCorpus experiments show limited zero-shot transfer, but do not remove this cold-start requirement. Finally, Jaccard overlap describes set-level agreement and cannot identify whether a disagreement reflects useful complementarity or semantic drift. These limitations motivate broader cross-LLM, multi-seed, and larger-query-set evaluation, but they do not affect the matched-input comparisons reported here.

\section{Conclusion}

We presented XViewRank, a lightweight conditional fusion method for LLM-augmented sparse retrieval. The method preserves normalized BM25 scores from the original, hypothesis, and step-back views, and uses result-list overlap to condition how those scores contribute to document ranking. Under matched inputs, XViewRank improves nDCG@10 and MAP numerically over fixed fusion on both TREC DL test years without jointly encoding query and passage text. The full-split statistical evidence is limited to DL19 MAP, and the smaller DL20 gains remain inconclusive. Our analyses further indicate that overlap is most useful as context when the views retrieve heterogeneous evidence, not as direct relevance evidence or as a reason to force query diversity. These results show that multi-view query expansion depends not only on generating useful views, but also on preserving and conditionally combining their retrieval evidence.

\appendix
\section{Reproducibility Details}
\label{app:reproducibility}

\subsection{Query-view generation prompts}

For every topic, the original query is used unchanged as the original view. The following instructions are supplied to GPT-4o mini, with the original query placed in the user message. Exactly one output is generated for each additional view and then cached.

\paragraph{Hypothesis view.}
\begin{quote}
\small
Write a concise passage that could answer the question and use terms likely to appear in relevant passages.
\end{quote}

\paragraph{Step-back view.}
\begin{quote}
\small
Rewrite the question into a higher-level retrieval query focusing on the underlying concept, mechanism, or broader information need.
\end{quote}

Generation uses a maximum of 500 output tokens, temperature 0.6, and top-\(p\) 0.9. Cached outputs are shared by XViewRank and all matched-view baselines so that fusion methods are compared on identical query text.

\subsection{Retrieval, training, and evaluation protocol}

\begin{table}[h]
\centering
\caption{Implementation and evaluation settings used for the main experiments.}
\label{tab:repro-settings}
\small
\setlength{\tabcolsep}{4pt}
\begin{tabular}{p{0.25\linewidth}p{0.66\linewidth}}
\toprule
Component & Setting \\
\midrule
Sparse retrieval & Pyserini BM25 over the MS MARCO passage index; \(k_1=0.9\), \(b=0.4\), depth 1,000 per view \\
Candidate construction & Union of the original, hypothesis, and step-back result lists, with duplicate document identifiers merged \\
Fusion features & Three per-query minimum-to-maximum normalized BM25 scores and three pairwise Jaccard overlaps \\
Fusion network & Two hidden layers of sizes 128 and 64, ReLU activations, and dropout 0.2 \\
Training pool & 6,317 queries and 9,040,208 pooled query-document candidates, including 6,602 positives \\
Optimization & AdamW; learning rate \(10^{-4}\); weight decay \(5\times10^{-4}\); batch size 1,024; at most 12 epochs \\
Deterministic execution & Fixed Python and CUDA seeds; sorted candidate identifiers; deterministic PyTorch and cuDNN operations; fixed cuBLAS workspace configuration \\
Model selection & Early-stopping patience 3 on the 2,000-query development split; positive-class weights capped at 200 \\
Test protocol & One development-selected configuration is applied unchanged to DL19 and DL20; test qrels are never used for selection \\
Evaluation & Graded qrels for nDCG; relevance grade at least 2 for MAP and Recall; cutoff 10 for nDCG and 1,000 for Recall \\
\bottomrule
\end{tabular}
\end{table}

The sensitivity and view-portfolio studies retrain the fusion scorer for each configuration. Consequently, their absolute scores can differ slightly from the fixed main run; conclusions in those studies are based on controlled comparisons within the same table. As an implementation-level reproducibility check, we independently repeated the complete seed-42 training and evaluation pipeline twice. The two runs produced byte-identical model checkpoints and identical DL19 and DL20 metrics. This establishes deterministic repeatability for a fixed seed; it is distinct from measuring variation across different training seeds. We will release cached views, split identifiers, view-specific retrieval runs, trained configurations, and evaluation scripts.

\section{Supplementary Neural-Reranker Reference}
\label{app:neural-reference}

This experiment positions XViewRank in a multi-stage retrieval pipeline without redefining the proposed method. The same MiniLM cross-encoder receives the original query and reranks the top 10 documents selected by either matched-view RRF or XViewRank; documents below this budget retain their input order. Table~\ref{tab:ce-reference} shows that semantic verification is complementary to retrieval-side fusion and that the XViewRank candidate order remains stronger under the same verifier and budget. The cross-encoder is not used to produce any main-table XViewRank result.

\begin{table}[h]
\centering
\caption{Supplementary neural-reranker reference with the same MiniLM cross-encoder and a verification budget of 10 documents.}
\label{tab:ce-reference}
\small
\setlength{\tabcolsep}{4pt}
\begin{tabular}{lcccc}
\toprule
Candidate order & \multicolumn{2}{c}{DL19} & \multicolumn{2}{c}{DL20} \\
\cmidrule(lr){2-3}\cmidrule(lr){4-5}
 & nDCG@10 & MAP & nDCG@10 & MAP \\
\midrule
RRF & 0.6376 & 0.4225 & 0.6152 & 0.4399 \\
XViewRank & 0.6583 & 0.4437 & 0.6250 & 0.4480 \\
CE on RRF top 10 & 0.6663 & 0.4448 & 0.6415 & 0.4540 \\
CE on XViewRank top 10 & \textbf{0.6763} & \textbf{0.4530} & \textbf{0.6533} & \textbf{0.4629} \\
\bottomrule
\end{tabular}
\end{table}

\bibliographystyle{splncs04}
\bibliography{references}

@inproceedings{cormack2009reciprocal,
  title={Reciprocal rank fusion outperforms condorcet and individual rank learning methods},
  author={Cormack, Gordon V and Clarke, Charles LA and Buettcher, Stefan},
  booktitle={Proceedings of the 32nd international ACM SIGIR conference on Research and development in information retrieval},
  pages={758--759},
  year={2009}
}

@article{zheng2023take,
  title={Take a step back: Evoking reasoning via abstraction in large language models},
  author={Zheng, Huaixiu Steven and Mishra, Swaroop and Chen, Xinyun and Cheng, Heng-Tze and Chi, Ed H and Le, Quoc V and Zhou, Denny},
  journal={arXiv preprint arXiv:2310.06117},
  year={2023}
}

@article{zhou2024hyqe,
  title={Hyqe: Ranking contexts with hypothetical query embeddings},
  author={Zhou, Weichao and Zhang, Jiaxin and Hasson, Hilaf and Singh, Anu and Li, Wenchao},
  journal={arXiv preprint arXiv:2410.15262},
  year={2024}
}

@inproceedings{xia2008listwise,
  title={Listwise approach to learning to rank: theory and algorithm},
  author={Xia, Fen and Liu, Tie-Yan and Wang, Jue and Zhang, Wensheng and Li, Hang},
  booktitle={Proceedings of the 25th international conference on Machine learning},
  pages={1192--1199},
  year={2008}
}

@inproceedings{cao2007learning,
  title={Learning to rank: from pairwise approach to listwise approach},
  author={Cao, Zhe and Qin, Tao and Liu, Tie-Yan and Tsai, Ming-Feng and Li, Hang},
  booktitle={Proceedings of the 24th international conference on Machine learning},
  pages={129--136},
  year={2007}
}

@article{craswell2020overview,
  title={Overview of the TREC 2019 Deep Learning Track},
  author={Craswell, Nick and Mitra, Bhaskar and Yilmaz, Emine and Campos, Daniel and Voorhees, Ellen M.},
  journal={arXiv preprint arXiv:2003.07820},
  year={2020}
}

@article{craswell2021overview,
  title={Overview of the TREC 2020 Deep Learning Track},
  author={Craswell, Nick and Mitra, Bhaskar and Yilmaz, Emine and Campos, Daniel},
  journal={arXiv preprint arXiv:2102.07662},
  year={2021}
}

@article{wang2023query2doc,
  title={Query2doc: Query expansion with large language models},
  author={Wang, Liang and Yang, Nan and Wei, Furu},
  journal={arXiv preprint arXiv:2303.07678},
  year={2023}
}

@inproceedings{liu2025exp4fuse,
  title={{Exp4Fuse}: A Rank Fusion Framework for Enhanced Sparse Retrieval using Large Language Model-based Query Expansion},
  author={Liu, Lingyuan and Zhang, Mengxiang},
  booktitle={Findings of the Association for Computational Linguistics: ACL 2025},
  pages={163--173},
  year={2025}
}

@inproceedings{gao2023precise,
  title={Precise zero-shot dense retrieval without relevance labels},
  author={Gao, Luyu and Ma, Xueguang and Lin, Jimmy and Callan, Jamie},
  booktitle={Proceedings of the 61st Annual Meeting of the Association for Computational Linguistics (Volume 1: Long Papers)},
  pages={1762--1777},
  year={2023}
}

@article{rocchio1971relevance,
  title={Relevance feedback in information retrieval},
  author={Rocchio Jr, Joseph John},
  journal={The SMART retrieval system: experiments in automatic document processing},
  year={1971},
  publisher={Englewood Cliffs}
}

@inproceedings{fox1994combination,
  title={Combination of Multiple Searches},
  author={Fox, Edward A. and Shaw, Joseph A.},
  booktitle={Proceedings of the 2nd Text REtrieval Conference (TREC-2)},
  pages={243--252},
  year={1994}
}

@inproceedings{lee1997analyses,
  title={Analyses of multiple evidence combination},
  author={Lee, Joon Ho},
  booktitle={Proceedings of the 20th annual international ACM SIGIR conference on Research and development in information retrieval},
  pages={267--276},
  year={1997}
}

@inproceedings{montague2002condorcet,
  title={Condorcet fusion for improved retrieval},
  author={Montague, Mark and Aslam, Javed A},
  booktitle={Proceedings of the eleventh international conference on Information and knowledge management},
  pages={538--548},
  year={2002}
}

@inproceedings{lavrenko2001relevance,
author = {Lavrenko, Victor and Croft, W. Bruce},
title = {Relevance based language models},
year = {2001},
isbn = {1581133316},
publisher = {Association for Computing Machinery},
address = {New York, NY, USA},
booktitle = {Proceedings of the 24th Annual International ACM SIGIR Conference on Research and Development in Information Retrieval},
pages = {120–127},
numpages = {8},
location = {New Orleans, Louisiana, USA},
series = {SIGIR '01}
}

@inproceedings{lillis2006probfuse,
  title={Probfuse: a probabilistic approach to data fusion},
  author={Lillis, David and Toolan, Fergus and Collier, Rem and Dunnion, John},
  booktitle={Proceedings of the 29th annual international ACM SIGIR conference on Research and development in information retrieval},
  pages={139--146},
  year={2006}
}

@article{robertson2009probabilistic,
  title={The probabilistic relevance framework: BM25 and beyond},
  author={Robertson, Stephen and Zaragoza, Hugo and others},
  journal={Foundations and Trends{\textregistered} in Information Retrieval},
  volume={3},
  number={4},
  pages={333--389},
  year={2009},
  publisher={Now Publishers, Inc.}
}

@inproceedings{aslam2001models,
  title={Models for metasearch},
  author={Aslam, Javed A and Montague, Mark},
  booktitle={Proceedings of the 24th annual international ACM SIGIR conference on Research and development in information retrieval},
  pages={276--284},
  year={2001}
}

@inproceedings{burges2005learning,
  title={Learning to rank using gradient descent},
  author={Burges, Chris and Shaked, Tal and Renshaw, Erin and Lazier, Ari and Deeds, Matt and Hamilton, Nicole and Hullender, Greg},
  booktitle={Proceedings of the 22nd international conference on Machine learning},
  pages={89--96},
  year={2005}
}

@inproceedings{lin2021pyserini,
  title={Pyserini: A Python toolkit for reproducible information retrieval research with sparse and dense representations},
  author={Lin, Jimmy and Ma, Xueguang and Lin, Sheng-Chieh and Yang, Jheng-Hong and Pradeep, Ronak and Nogueira, Rodrigo},
  booktitle={Proceedings of the 44th international ACM SIGIR conference on research and development in information retrieval},
  pages={2356--2362},
  year={2021}
}

@article{larsen2009data,
  title={Data fusion according to the principle of polyrepresentation},
  author={Larsen, Birger and Ingwersen, Peter and Lund, Berit},
  journal={Journal of the American Society for Information Science and Technology},
  volume={60},
  number={4},
  pages={646--654},
  year={2009},
  publisher={Wiley Online Library}
}

@article{nogueira2019doc2query,
  title={From doc2query to {docTTTTTquery}},
  author={Nogueira, Rodrigo and Lin, Jimmy},
  journal={Online Preprint},
  volume={6},
  number={2},
  pages={2},
  year={2019}
}

@inproceedings{shokouhi2007segmentation,
  title={Segmentation of search engine results for effective data-fusion},
  author={Shokouhi, Milad},
  booktitle={European Conference on Information Retrieval},
  pages={185--197},
  year={2007},
  organization={Springer}
}

@article{jagerman2023query,
  title={Query expansion by prompting large language models},
  author={Jagerman, Rolf and Zhuang, Honglei and Qin, Zhen and Wang, Xuanhui and Bendersky, Michael},
  journal={arXiv preprint arXiv:2305.03653},
  year={2023}
}

@article{furnas1987vocabulary,
  title={The vocabulary problem in human-system communication},
  author={Furnas, George W. and Landauer, Thomas K. and Gomez, Louis M. and Dumais, Susan T.},
  journal={Communications of the ACM},
  volume={30},
  number={11},
  pages={964--971},
  year={1987},
  publisher={ACM New York, NY, USA}
}

@article{li2026query,
  title={Query Expansion in the Age of Pre-trained and Large Language Models: A Comprehensive Survey},
  author={Li, Minghan and Lv, Xinxuan and Zou, Junjie and Chen, Tongna and Zhang, Chao and An, Suchao and Nie, Ercong and Zhou, Guodong},
  journal={ACM Transactions on Information Systems},
  volume={44},
  number={5},
  pages={1--42},
  year={2026},
  publisher={ACM New York, NY}
}

@inproceedings{li2026s2g,
  title={S2G-RAG: Structured Sufficiency and Gap Judging for Iterative Retrieval-Augmented QA},
  author={Li, Minghan and Zou, Junjie and Lv, Xinxuan and Zhang, Chao and Zhou, Guodong},
  booktitle={Proceedings of the 64th Annual Meeting of the Association for Computational Linguistics (Volume 1: Long Papers)},
  pages={25846--25862},
  year={2026}
}

@article{li2025survey,
  title={A survey of long-document retrieval in the PLM and LLM era},
  author={Li, Minghan and Zhang, Yishuai and Lv, Tianrui and Zhao, Siqi and Luo, Miyang and Nie, Ercong and Zhou, Guodong},
  journal={ACM Transactions on Information Systems},
  year={2025},
  publisher={ACM New York, NY}
}

@article{li2026dual,
  title={Dual-layer prompt ensembles: Leveraging system-and user-level instructions for robust LLM-based query expansion and rank fusion},
  author={Li, Minghan and Nie, Ercong and Huang, Huiping and Lv, Xinxuan and Zhou, Guodong},
  journal={Information Fusion},
  pages={104160},
  year={2026},
  publisher={Elsevier}
}

@inproceedings{li2026glier,
  title={GLIER: Generative Legal Inference and Evidence Ranking for Legal Case Retrieval},
  author={Li, Minghan and Lv, Tianrui and Zhang, Chao and Zhou, Guodong},
  booktitle={Proceedings of the 64th Annual Meeting of the Association for Computational Linguistics (Volume 1: Long Papers)},
  pages={29561--29572},
  year={2026}
}

\end{document}